%% file: main.tex
\documentclass[nofootinbib,aps,prl,reprint,amsmath,amssymb,groupedaddress]{revtex4-2}

\usepackage{graphicx}
\usepackage{hyperref}
\usepackage[acronym,nomain,hyperfirst=false]{glossaries}
\usepackage[capitalise]{cleveref}
\usepackage{bm,braket,xfrac}
\usepackage{booktabs,tabularx,dcolumn,siunitx}

\input{math_macros}
\newcommand{\exciting}{\texttt{exciting}}
\newcommand{\sub}[1]{\textsubscript{#1}}
\DeclareSIUnit\angstrom{\protect \text {Å}}
\newcommand{\ie}{{\it i.e.}, }

\newcommand{\eg}{{\it e.g.}, }
\newcommand{\wrt}{with respect to }

\begin{document}
\glsdisablehyper 

\title{Automatic Generation of Maximally Localized Wannier Functions via Optimized Projection Functions and Self-projections}

\author{Sebastian Tillack}
\email{sebastian.tillack@physik.hu-berlin.de}
\author{Claudia Draxl}
\affiliation{Institut f\"ur Physik and CSMB, Humboldt-Universit\"at zu Berlin, Berlin, Germany}

\date{\today}
\input{acronyms}

\begin{abstract}
We present an automatized approach towards \glspl{mlwf} applicable to both occupied and unoccupied states. We overcome limitations of the standard \gls{opf} method and its approximations by providing an exact expression for the gradient of the Wannier spread functional \wrt a single semi-unitary \gls{opf} matrix. Moreover, we demonstrate that the localization of the resulting \glspl{wf} can be further improved by including projections on reasonably localized \glspl{wf}, so-called self-projections.
\end{abstract}
\glsresetall

\maketitle

Within the last two decades, \glspl{mlwf} have become a powerful tool in theoretical solid-state physics, with widespread applications in chemical analysis \cite{Abu-Farsakh2007}, electric polarization, magnetism \cite{Wang2006}, electron--phonon coupling \cite{Giustino2007}, charge and heat transport \cite{Protik2022}, superconductivity \cite{Kempa2023}, and more. In general, \glspl{wf} are defined by a Fourier-like transform of Bloch functions,
\begin{align}
    \label{eq:wf_def}
    w_{n\vec{R}}(\vec{r}) &= \frac{1}{N_{\vec{k}}} \sum_{\vec{k}} \E^{-\I \vec{k} \cdot \vec{R}} 
    \sum_{m=1}^{J_{\vec{k}}} U_{mn}^{\vec{k}}\, \psi_{m\vec{k}}(\vec{r}) \;,
\end{align}
with a set of (semi-)unitary matrices $\lbrace \mat{U}^{\vec{k}} \rbrace$, called a gauge. Finding \glspl{mlwf} amounts to finding an optimal gauge that minimizes the spatial spread of the \glspl{wf}. Commonly, the spread introduced by \citet{Marzari1997}
\begin{align}
    \label{eq:spread_functional}
    \Omega [\lbrace \mat{U}^{\vec{k}} \rbrace] &= \sum_{n=1}^J \left[ \braket{w_{n\vec{0}} | r^2 | w_{n\vec{0}}} - \braket{w_{n\vec{0}} | \vec{r} | w_{n\vec{0}}}^2 \right]
\end{align}
is employed. However, recently, an alternative localization measure was proposed \cite{Li2024}. The spread is minimized using gradient-based methods such as steepest descent \cite{Marzari1997,Damle2019}. The parameter space of this optimization problem grows rapidly with the number of $\vec{k}$ vectors and bands involved, increasing the chance of converging to a false local minimum or not converging at all. Providing a good starting point for the minimization, \ie an initial gauge $\lbrace \mat{U}^{\vec{k}} \rbrace$, is indispensable for achieving maximal localization. The standard approach is based on the projection
\begin{align}
    \label{eq:A_def}
    A^{\vec{k}}_{mn} = \braket{ \psi_{m\vec{k}} | g_n }
\end{align}
of the Bloch states on a set of projection functions $g_n(\vec{r})$, which should approximate the $J$ desired \glspl{mlwf} in the considered unit cell, $w_{n\vec{0}}(\vec{r})$. The initial gauge is set to the unitary matrices closest to $\mat{A}^{\vec{k}}$, which can be obtained from the \gls{svd} of $\mat{A}^{\vec{k}}$ via
\begin{align}
    \label{eq:U_A_def}
    \mat{U} = \mat{U}_{\mat{A}} &\equiv \mat{V}\, \mat{W}^\dagger \; &\mat{A} &= \mat{V}\, \mat{\Sigma}\, \mat{W}^\dagger \;,
\end{align}
where we dropped the superscript $\vec{k}$ for brevity. The projection functions $g_n(\vec{r})$ have to be chosen manually based on chemical intuition and additional knowledge about the chemical bonding in the considered material. This is a serious limitation for complex materials whose chemical characteristics are yet to be investigated. The same holds for the calculation of \glspl{wf} describing delocalized (unoccupied) states away from the Fermi level, which don't form simple (anti) bonding states. In such cases, the random generation of an initial gauge may be a last resort. Most importantly, however, not having a good initial guess impedes full automatized calculations of \glspl{mlwf}, which are desirable as computational material science and design has entered the realm of high-throughput calculations and data-centric approaches.

In this Letter, we address the challenge of providing suitable projection functions by reformulating it as a mathematical optimization problem and providing an algorithm for its solution. The success of the method is demonstrated by the calculation of \glspl{mlwf} corresponding to both isolated (occupied) and entangled (unoccupied) bands in eight different materials.

Our approach extends the \gls{opf} method \cite{Mustafa2015}. The underlying idea is to expand the $J$ projection functions $g_n(\vec{r})$ within a larger set of $M \geq J$ orthonormal trial orbitals $h_j(\vec{r})$ as
\begin{align}
    \label{eq:opf_def}
    g_n(\vec{r}) &= \sum_{j=1}^M X_{jn}\, h_j({\vec{r}}) \;,
\end{align}
with a semi-unitary coefficient matrix $\mat{X}$. The gauge $\lbrace \mat{U}^{\vec{k}} \rbrace$ is obtained from \cref{eq:A_def,eq:U_A_def} by setting ${\mat{A} = \braket{\psi_{\vec{k}}|h}}$ and $\mat{U} = \mat{U}_{\mat{AX}}$. The optimal $\mat{X}$ is to be found by minimizing the spread $\Omega$ that implicitly depends on it. The advantage is that $\Omega$ is minimized \wrt a single matrix $\mat{X}$ instead of a set of $N_{\vec{k}}$ matrices $\lbrace \mat{U}^{\vec{k}} \rbrace$. This vastly reduces the parameter space of the optimization problem and thus the computational cost, simultaneously facilitating more stable convergence. Setting up the pool of trial orbitals $h_j(\vec{r})$ is far less restrictive than directly selecting the projection functions $g_n(\vec{r})$. The trial orbitals should approximately span the space of the desired \glspl{mlwf}. This approach leads directly to \glspl{mlwf}, if $\operatorname{span}\lbrace h_j \rbrace = \operatorname{span}\lbrace w_{n\vec{0}} \rbrace$. The manual selection of projection functions gets transformed into the automatic determination of the \gls{opf} matrix $\mat{X}$, encoding the chemical characteristics of the projection functions, which were previously a necessary user input. On the other hand, minimizing $\Omega$ \wrt $\mat{X}$ is difficult due to the complex nonlinear dependence determined by \cref{eq:U_A_def}. \citet{Mustafa2015} simplified the problem introducing two major approximations: (i) The problem is linearized assuming $\mat{U}_{\mat{AX}} = \mat{U}_{\mat{A}}\,\mat{X}$, and an auxiliary Lagrangian $\mathcal{L}$ is minimized instead of $\Omega$, accounting for the constraint of $\mat{AX}$ being unitary (justifying the previous assumption) by a Lagrangian multiplier $\lambda$. (ii) Only the gauge-independent and off-diagonal part of the spread functional, $\Omega_\textrm{I,OD}$, is minimized, neglecting the diagonal part, $\Omega_\textrm{D}$ (see \cite{Marzari1997} for the definitions of the individual parts of the spread functional). Approximation (ii) is justified close to the minimum of $\Omega$, where typically $\Omega_{\rm D} \ll \Omega_{\rm I,OD}$. Approximation (i) introduces another parameter, $\lambda$, which determines the success of the method and is generally case dependent.

Here we provide an expression for the gradient of the spread functional $\Omega$ \wrt the \gls{opf} matrix $\mat{X}$, which does not rely on any approximations and can be used in a gradient-based optimization algorithm on the manifold of unitary matrices (Stiefel manifold). Our derivations are based on the differentiation of the \gls{svd} in \cref{eq:U_A_def}, \ie calculating the variation of the singular values ${\D\mat{\Sigma}}$ and the left and right singular vectors ${\D\mat{V}}$ and ${\D\mat{W}}$ upon a variation ${\D(\mat{AX}) = \mat{A}\, \D\mat{X}}$. This allows us to compute the variation of the (semi-)unitary rotations $\mat{U}$ as ${\D\mat{U} = \D\mat{V}\, \mat{W}^\dagger + \mat{V}\, \D\mat{W}^\dagger}$ and hence the derivative of $\mat{U}$ \wrt $\mat{X}$. The derivative of the spread functional can then be obtained by applying the chain rule for matrix derivatives. After extensive algebra (see Supplementary Material), we arrive at the following expression for the gradient of the spread functional \wrt the \gls{opf} matrix $\mat{X}$, which constitutes the main result of this work:
\begin{subequations}
    \label{eq:omega_grad_x}
\begin{align}
    \label{eq:omega_grad_x_def}
    \mGrad{\Omega}{X} &= \sum_{\vec{k}}\, \mat{A}^\dagger\, \left\lbrace 
    \mat{V}\, \left( \mat{F} \odot \left[ \mat{V}^\dagger\, \mGrad{\Omega}{U}\, \mat{W} \right] - \text{H.c.} \right)\, \mat{W}^\dagger \right. \nonumber \\
    &\quad + (\mat{I}_{J_{\vec{k}}} - \mat{V}\, \mat{V}^\dagger) \mGrad{\Omega}{U}\, \mat{W}\, \mat{\Sigma}^{-1}\, \mat{W}^\dagger \nonumber \\
    &\quad + \left. \mat{V}\, \mat{\Sigma}^{-1}\, \mat{V}^\dagger\, \mGrad{\Omega}{U}\, (\mat{I}_J - \mat{W}\, \mat{W}^\dagger) \right\rbrace \;, 
\end{align}
\begin{align}
    \label{eq:omega_grad_x_supplement}
    \Sigma_{ij} &= \sigma_i\, \delta_{ij} \;, &
    F_{ij} &= \begin{cases} \frac{1}{2\sigma_i} & \sigma_i = \sigma_j \\ \frac{\sigma_j - \sigma_i}{\sigma_j^2 - \sigma_i^2} & \sigma_i \neq \sigma_j \end{cases} \;.
\end{align}
\end{subequations}
Here $\odot$ denotes the (element-wise) Hadamard product, $\mGrad{\Omega}{U}$ is the Euclidean gradient of the spread functional \wrt the unitary rotations $\mat{U}$ as provided by \citet{Damle2019} (which is different from the gradient given in \cite{Marzari1997}). $\mat{\Sigma}$, $\mat{V}$, and $\mat{W}$ are defined via \cref{eq:U_A_def} by replacing $\mat{A}$ with $\mat{AX}$, and $\mat{I}_J$ is the ${J \times J}$ identity matrix. 

There are a few comments to be made on \cref{eq:omega_grad_x}: (i) All matrices in \cref{eq:omega_grad_x} except for the \gls{opf} matrix $\mat{X}$ carry an implicit superscript $\vec{k}$, which has been dropped for brevity. (ii) The third line in \cref{eq:omega_grad_x_def} vanishes if $\mat{AX}$ is not singular, \ie $\mat{W}$ is square. This is always the case if all trial orbitals $h_j(\vec{r})$ are linearly independent, especially if they are constructed as outlined in the following. (iii) If, in addition, the Bloch states $\psi_{m\vec{k}}$ span a $J$-dimensional subspace, \ie in the case of an isolated group of bands \cite{Marzari1997} or after the disentanglement step \cite{Souza2001}, then also the second line in \cref{eq:omega_grad_x_def} vanishes.

The second ingredient (besides the gradient expression) required to successfully generate \glspl{opf} is a suitable set of trial orbitals $h_j(\vec{r})$. We wish this set to be well localized, to approximately span the space of the \glspl{mlwf}, and to be as small as possible. We approach these requirements by using linear combinations of localized atom-centered orbitals $\tilde{h}_k(\vec{r})$ as trial orbitals,
\begin{align}
    \label{eq:trial_def}
    h_j(\vec{r}) &= \sum_k B_{kj}\, \tilde{h}_k(\vec{r}) \;.
\end{align}
The coefficients $\mat{B}$ are determined such that the trial orbitals are orthonormal and have maximal overlap with the subspace of Bloch states that is to be wannierized. This idea is based on the observation that the \glspl{mlwf} $w_{n\vec{0}}(\vec{r})$ are given by linear combinations of the Bloch states at all $\vec{k}$-points within the considered subspace. We maximize the overlap by solving the generalized eigenvalue problem
\begin{subequations}
    \label{eq:gep_trial}
\begin{align}
    \mat{P}\, \mat{B} &= \mat{S}\, \mat{B}\, \mat{\Lambda} \;,
\end{align}
with
\begin{align}
    P_{ij} &= \braket{\tilde{h}_i | \hat{\mat{P}} | \tilde{h}_j} = \frac{1}{N_{\vec{k}}} \sum_{\vec{k}} \sum_{m=1}^J \braket{\tilde{h}_i | \psi_{m\vec{k}}} \braket{\psi_{m\vec{k}} | \tilde{h}_j } \;, \\
    S_{ij} &= \braket{\tilde{h}_i | \tilde{h}_j} \;,
\end{align}
\end{subequations}
and setting the coefficients $B_{kj}$ in \cref{eq:trial_def} to the eigenvectors corresponding to the $M$ largest eigenvalues $\lambda_j$. The normalized sum $\gamma = J^{-1} \sum_j \lambda_j$ of the $M$ largest eigenvalues is a measure of the subspace coverage. If ${\gamma = 1}$, then the trial orbitals $h_j(\vec{r})$ fully span the $J$-dimensional subspace of the Bloch states $\psi_{m\vec{k}}(\vec{r})$. In practice, one can either fix a value for $M$ and take the $M$ largest eigenvalues, or $M$ is determined by taking all eigenvalues larger than a fixed threshold. In this work, we employ the second choice. Note that the coefficients $\mat{B}$ only need to be calculated once at the beginning. We then minimize the spread using a fixed set of $M$ trial orbitals $h_j(\vec{r})$. A very similar approach is used to initialize the \gls{opf} matrix $\mat{X}$. At the beginning of the iterative minimization of $\Omega$, we set $\mat{X}$ to the eigenvectors of $\braket{h | \hat{\mat{P}} | h}$ corresponding to the $J$ largest eigenvalues.

The proposed algorithm has been implemented in the full-potential all-electron code \exciting, which employs a \gls{lapw} and \gls{lo} basis \cite{Gulans2014}. We use \glspl{lo} as atomic projectors $\tilde{h}_k(\vec{r})$ \cite{Tillack2020}, which are given by products of radial functions and spherical harmonics and are confined within a sphere around the nuclei. The radial functions are solutions of a radial Schr\"odinger or Dirac equation with a spherically symmetric potential. For each atomic species, we automatically compute such radial functions up to a given principal quantum number $n$ (determining the number of nodes in the radial function) and construct the corresponding \glspl{lo} according to the aufbau principle. We expect that other, \eg analytic radial functions such as spherical Bessel functions or Gaussians, will work as well. For the minimization of the spread functional, we employ the \gls{lbfgs} algorithm on the manifold of unitary matrices as implemented in \cite{Boumal2014}.    

We apply our algorithm to six different materials, adopting the list of materials and numerical settings used in Ref. \cite{Mustafa2015}, and adding the oxide perovskite BaSnO\sub{3}. In all cases, trial orbitals corresponding to eigenvalues ${\lambda > 0.01}$ in \cref{eq:gep_trial} are included. The results are summarized in \cref{tab:spread-isolated}.
\begin{table}
    \caption{\label{tab:spread-isolated} \gls{wf} spread (in \si{\angstrom^2}) from \glspl{opf}, $\Omega^\textrm{OPF}$, and \glspl{mlwf} (after subsequent minimization), $\Omega^\textrm{MLWF}$, for valence bands in different materials. In all cases, the $J$ highest bands below the Fermi level were wannierized. $M$ is the total number of trial orbitals used to construct the \glspl{opf} and $\gamma$ their respective subspace coverage. Atomic projectors were either restricted to the home unit cell or include additional copies from nearest neighbor atoms in neighboring cells (+nn). {Si-20} refers to strongly distorted silicon with 20 atoms in the primitive cell.}
    \input{tables/spread_isolated}
\end{table}
For all materials studied, the spread of the \glspl{wf} obtained from the \gls{opf} method is within $2\%$ deviation from the \glspl{mlwf} that are obtained from a full minimization of the spread functional in the space of ${\lbrace \mat{U}^{\vec{k}} \rbrace}$, using the \gls{opf} result as starting point. By including identical copies of the atomic trial orbitals centered at nearest-neighbor atoms and thus accounting for all possible bonds in the system, this deviation can be further reduced to less than $1\%$. We stress that a linear combination of atom-centered trial orbitals is also capable of producing bond-centered \glspl{wf}, as in the case of silicon.

In the case of entangled bands, we first disentangle an optimal $J$-dimensional subspace using an inner (frozen) and outer energy window as described in \cite{Souza2001}. As examples, we consider the lowest conduction bands in \mbox{Si-20} and SiO\sub{2} as well as zinc as a metallic system. We choose wide energy windows (up to 85\,\si{eV}) from which we generate a large number of \glspl{wf} (up to 160). The corresponding results are summarized in \cref{tab:spread-entangled}. In contrast to the isolated-band case, the \gls{opf} spread is further away from the \glspl{mlwf}. Hence, here, the \gls{opf} approach is not meant to be used as a stand-alone method. However, it provides a very good starting point for a subsequent minimization of the spread. In the presented examples, the inclusion of nearest-neighbor atoms yields a substantial gain in the \gls{opf} spread. This is due to the increase in the number of trial orbitals, $M$, and hence the variational freedom in finding the \glspl{opf}. Note that for \mbox{Si-20} the subsequent minimization converges to exactly the same spread $\Omega^{\rm MLWF}$ starting from both \glspl{opf} with and without nearest neighbor inclusion, while for SiO\sub{2} and Zn it yields almost, but not exactly, the same spread in both cases.

\begin{table}
    \caption{\label{tab:spread-entangled} Same as \cref{tab:spread-isolated}, but for entangled bands. $J$ \glspl{wf} were disentangled from the inner (outer) energy window $E^{\rm wind}$ (in \si{eV} above the Fermi level).}
    \input{tables/spread_entangled}
\end{table}
\begin{figure}
    \includegraphics[width=\linewidth]{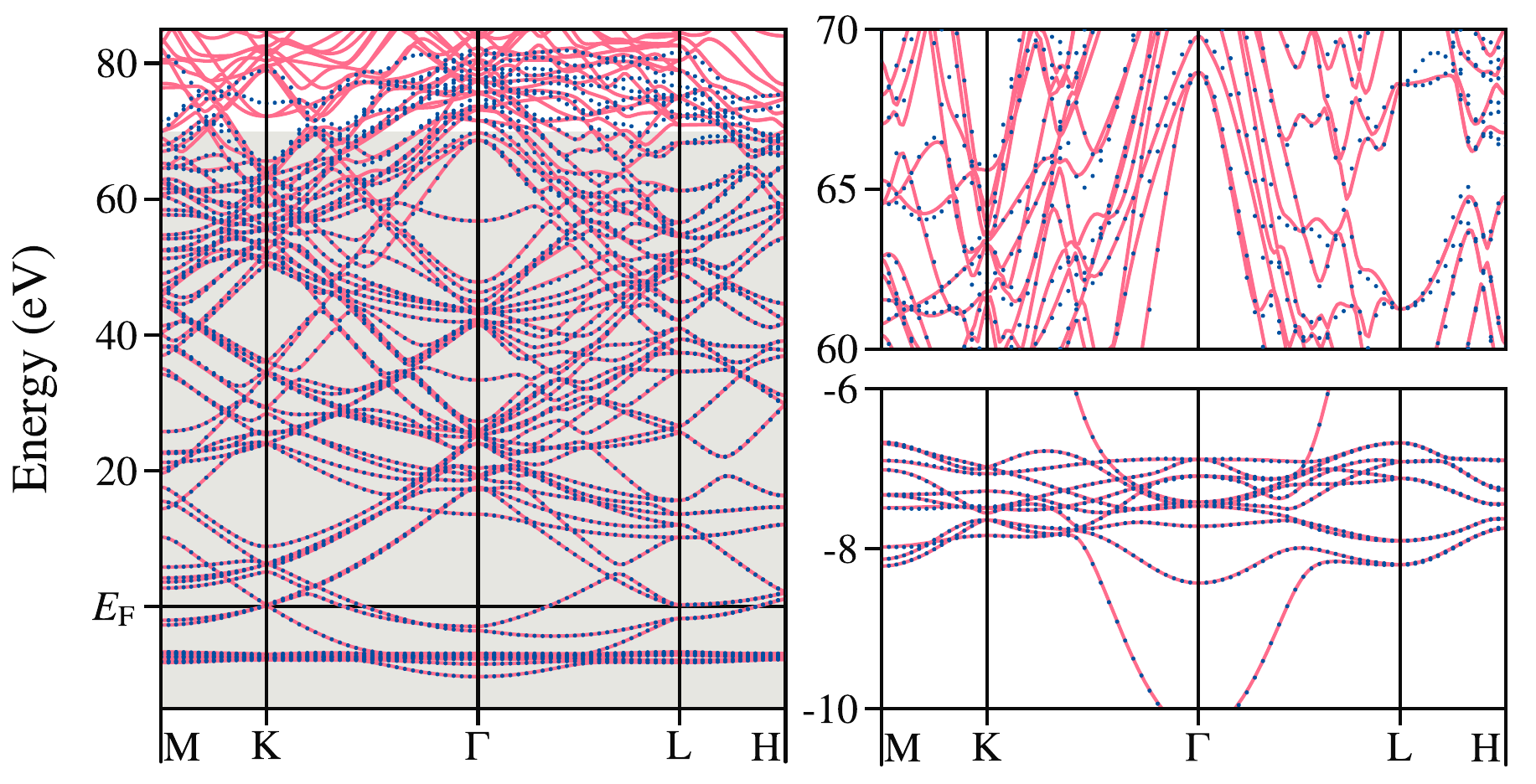}
    \caption{\label{fig:bandstructure-Zn} Bandstructure of Zn calculated by direct diagonalization of the \gls{dft} Hamiltonian (red lines) and Wannier interpolation (blue dots). The full energy range displayed in the left panel coincides with the choice of the outer energy window. The inner window is illustrated by the gray background. Top right: Detailed view of the upper part of the inner energy window. Bottom right: Detailed view of the ten flat $d$-like bands.}
\end{figure}

\begin{figure}
    \includegraphics[width=0.9\linewidth]{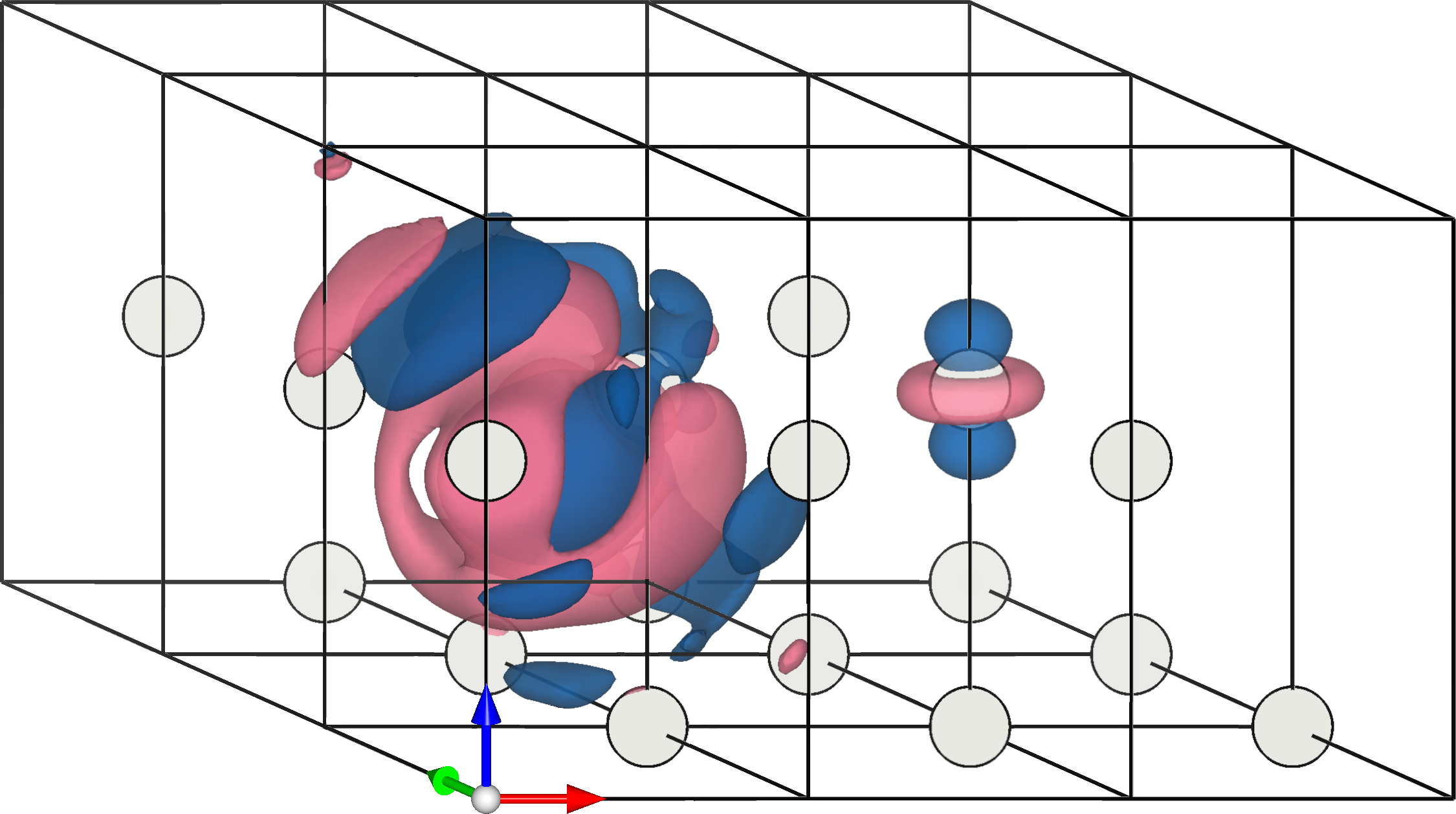}
    \caption{\label{fig:wf-Zn} Illustration of two \glspl{mlwf} in Zn. The small \gls{wf} on the right is one of ten strongly localized \glspl{wf} associated with the flat $d$-like bands. The \gls{wf} on the left is randomly selected. The displayed iso-surface encloses $90\,\%$ of the total charge.}
\end{figure}

In \cref{fig:bandstructure-Zn} we compare the bandstructure of Zn obtained from a direct diagonalization of the \gls{dft} Hamiltonian with the bandstructure obtained from Wannier interpolation. There is excellent agreement up to the upper bound of the inner energy window at 70\,\si{eV}. Also the ten flat $d$ bands in the occupied region are perfectly reproduced. They are represented by ten out of the 64 \glspl{wf}, which are strongly localized with a spread of only 0.2\,\si{\angstrom^2}. One of them is illustrated on the right in \cref{fig:wf-Zn} along with one of the remaining 54 \glspl{wf}. They are both real-valued, indicating that they are truly maximally localized.

The larger difference between $\Omega^{\rm OPF}$ and $\Omega^{\rm MLWF}$ in the case of entangled bands originates from the incompleteness of the pool of trial orbitals $h_j(\vec{r})$, \ie it does not fully span the space of the \glspl{mlwf}. One idea to improve upon this incompleteness is to include $J$ reasonably (but not necessarily maximally) localized \glspl{wf} $w_{n\vec{0}}(\vec{r})$ in an extended set of $M+J$ trial orbitals
\begin{align}
    \label{eq:trial_sp}
    \begin{Bmatrix}\everymath{\displaystyle}\begin{array}{rll}
    h^\textrm{sp}_j(\vec{r}) &= h_j(\vec{r}) \phantom{\sum_i^M}\\
    h^\textrm{sp}_{M+n}(\vec{r}) &= \sum_{m=1}^J C_{mn}\, w_{m\vec{0}}(\vec{r}) + \sum_{i=1}^M D_{in}\, h_i(\vec{r})
    \end{array}\end{Bmatrix} \; , 
\end{align}
with $j=1,\dots,M$ and $n=1,\dots,J$. The coefficients $C_{mn}$ and $D_{in}$ can be chosen such that the extended set $h^\textrm{sp}_j(\vec{r})$ is still orthonormal. Since we choose \glspl{wf} themselves to be part of the pool of trial orbitals, we call this method self-projection. The overlap between the Bloch states and this extended set of trial orbitals reads
\begin{align}
    \label{eq:A_trial_sp_def}
    \mat{A}^{\vec{k},\textrm{sp}} &= \begin{bmatrix} \mat{A}^{\vec{k}}, &
    \mat{U}^{\vec{k}}\, \mat{C} + \mat{A}^{\vec{k}}\, \mat{D} \end{bmatrix} &\in \mathbb{C}^{J \times (M+J)} \;,
\end{align}
where ${\lbrace \mat{U}^{\vec{k}} \rbrace}$ describes a fixed initial gauge that defines the \glspl{wf} in \cref{eq:trial_sp} via \cref{eq:wf_def}. We aim to find $J$ \glspl{opf} $g_n^\textrm{sp}(\vec{r})$ given by the extended \gls{opf} matrix $\mat{X}^\textrm{sp}$ via \cref{eq:opf_def}. Due to our primary assumption of an initial gauge that defines already reasonably localized \glspl{wf} $w_{n\vec{0}}(\vec{r})$, we can choose these $w_{n\vec{0}}(\vec{r})$ as an initial guess for the \glspl{opf} and initialize $\mat{X}^\textrm{sp}$ accordingly. We optimize $\mat{X}^\textrm{sp}$ using the \gls{lbfgs} algorithm and find an improved gauge ${\lbrace \mat{U}^{\vec{k}} \rbrace}$ from ${\mat{U} = \mat{U}_{\mat{AX}^{\rm sp}}}$. This improved gauge is then used to update the \glspl{wf} in \cref{eq:trial_sp}, and the procedure is repeated iteratively. Note that the spread functional $\Omega$ is invariant under a unitary mixing of the \glspl{wf}. Therefore, it is not possible to use solely \glspl{wf} as trial orbitals, because every \gls{opf} matrix $\mat{X}$ will lead to the same value for $\Omega$.

\begin{figure}
    \includegraphics[width=\linewidth]{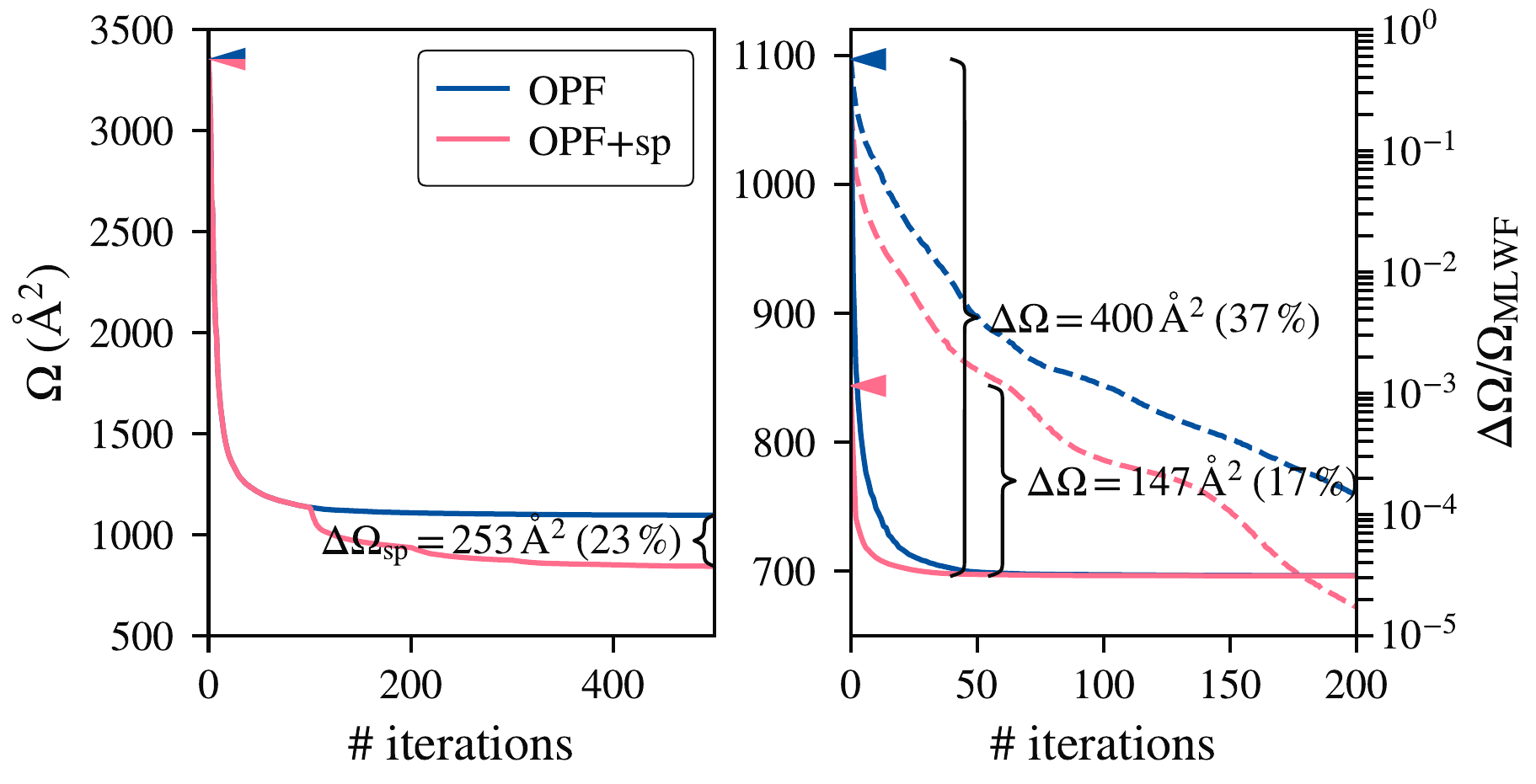}
    \caption{\label{fig:convergence-Si20} Left: Reduction of the spread $\Omega$ in the \gls{opf} step without (dark blue) and with (light red) self-projection with a self-projection cycle of 100 iterations. Right: Reduction of the spread $\Omega$ in the \gls{mlwf} step (solid lines, left axis) and relative error in the final \gls{mlwf} spread (dashed lines, right axis). Initial spreads are marked by arrows on the left axes.}
\end{figure}

We employ the iterative self-projection scheme in cases of entangled bands with a self-projection cycle of 100 iterations, \ie we run 100 optimization steps for $\mat{X}$ to find a gauge corresponding to reasonably localized \glspl{wf}. We now include these \glspl{wf} in the extended set of trial orbitals $h_j^{\rm sp}(\vec{r})$. The extended \gls{opf} matrix $\mat{X}^{\rm sp}$ is optimized for another 100 iterations producing an improved set of \glspl{wf}, which replace the previous ones in the extended set of trial orbitals. The last step is repeated four times amounting to 500 iterations in total (100 initial iterations + 4 self-projection cycles \`a 100 iterations).
\begin{table}
    \caption{\label{tab:spread-self-projection} Same parameters as in \cref{tab:spread-entangled} but employing self-projection. $\Omega^{\rm OPF+sp}$ is the spread from \glspl{opf} using the self-projection scheme. $\Delta\Omega^{\rm sp}$ is the localization gain compared to \gls{opf} without self-projection.}
    \input{tables/spread_self-projection}
\end{table}
By this, we are able to achieve an additional localization gain of more than 20\% over the \gls{opf} approach without self-projection for both \mbox{Si-20} and SiO\sub{2} (see \cref{tab:spread-self-projection}). For Zn, in contrast, the gain is substantially smaller, \ie the inclusion of intermediate \glspl{wf} does not significantly increase the span of the trial orbitals. This may be due to the atomic-like nature of the \glspl{mlwf} in Zn, which are already similar to the atomic trial orbitals. In \mbox{Si-20}, the \gls{opf} spread without self-projection is 1097.7\,\si{\angstrom^2} compared to 835.9\,\si{\angstrom^2} with self-projection. The latter provides an improved starting point for the calculation of \glspl{mlwf}, allowing for a more rapid convergence of the minimization procedure as shown in \cref{fig:convergence-Si20}.

In conclusion, we have demonstrated how to eliminate the error-prone task of manually providing projection functions to initialize the search for \glspl{mlwf}. This is achieved by a gradient-based algorithm to compute \glspl{opf} from automatically generated trial orbitals. Our scheme does not require any additional user input beyond the energy range to be wannierized. The calculation of \glspl{wf} from the valence bands in different materials demonstrate that our approach is able to essentially reach maximal localization for isolated bands. For entangled bands, it produces an excellent starting point for the calculation of \glspl{mlwf}, which can be further improved by self-projections. Thus, our work provides an additional and essential building block towards the fully automatized generation of \glspl{mlwf} in high-throughput calculations. In the future, we aim to combine our method with existing approaches by using alternative trial functions such as \glspl{pao} \cite{Qiao2023a} and \gls{scdm} \cite{Damle2017,Damle2018}.

\bibliography{bibliography}

\end{document}


\maketitle
\renewcommand{\thesection}{S\arabic{section}}
\sloppy

\section{Definitions and Useful Manipulations}
We follow the notation used in \citet{Hjorungnes2011}. The \textit{Kroneker product} (vector product) of two matrices $\mat{A}$ and $\mat{B}$ is written as ${\mat{A} \otimes \mat{B}}$. The \textit{Hadamard product} (elemenwise product) of two matrices $\mat{A}$ and $\mat{B}$ is written as ${\mat{A} \odot \mat{B}}$. The \textit{vectorization} of a matrix $\mat{A}$ is written as $\mvec{\mat{A}}$ and represents the vector obtained by stacking the columns of $\mat{A}$ in the normal order to a single column vector. The \textit{diagonal matrix} containing the elements of vectors $\vec{a}$ on the diagonal is written as $\diag{\vec{a}}$. \textit{Differentials} of scalar, vector, or matrix functions are written as ${\D f}$, ${\D \vec{f}}$, or ${\D \mat{F}}$, respecively. The \textit{derivative} of a complex valued matrix function ${\mat{F}(\mat{Z},\mat{Z}^\ast) : \mathbb{C}^{M \times N} \times \mathbb{C}^{M \times N} \rightarrow \mathbb{C}^{P \times Q}}$ \wrt $\mat{Z}$ and $\mat{Z}^\ast$ are denoted $\mderiv{\mat{Z}}{\mat{F}}$ and $\mderiv{\mat{Z}^\ast}{\mat{F}}$, respectively, and are implicitly defined by
\begin{equation}
    \label{eq:mderiv_def}
    \mvec{\D\mat{F}} = \mderiv{\mat{Z}}{\mat{F}}\, \mvec{\D\mat{Z}} + \mderiv{\mat{Z}^\ast}{\mat{F}}\, \mvec{\D\mat{Z}^\ast}
\end{equation}
and $\mderiv{\mat{Z}}{\mat{F}}$ is a complex matrix of form ${PQ \times MN}$. A useful property of the derivative is ${\mderiv{\mat{Z}^\ast}{\mat{F}^\ast} = \mderiv{\mat{Z}}{\mat{F}}^\ast}$. With this definition of the derivative, the \textit{chain rule} for a composed function ${\mat{F}(\mat{Z},\mat{Z}^\ast) = \mat{G}(\mat{H}(\mat{Z},\mat{Z}^\ast), \mat{H}^\ast(\mat{Z},\mat{Z}^\ast))}$ is given by
\begin{equation}
    \label{eq:chain-rule}
    \mderiv{\mat{Z}}{\mat{F}} = \mderiv{\mat{H}}{\mat{G}} \mderiv{\mat{Z}}{\mat{H}} + \mderiv{\mat{H}^\ast}{\mat{G}} \mderiv{\mat{Z}}{\mat{H}^\ast} \;.
\end{equation}
The \textit{gradient} of a real scalar function ${f(\mat{Z},\mat{Z}^\ast) : \mathbb{C}^{M \times N} \times \mathbb{C}^{M \times N} \rightarrow \mathbb{R}}$ is a complex ${M \times N}$ matrix which is defined as
\begin{equation}
    \mgrad{\mat{Z}}{f}_{ij} = \frac{\partial f}{\partial z_{ij}^\ast} \quad;\; i=1,\ldots,M\;;\;j=1,\ldots,N\;. 
\end{equation}
This defines the direction of maximum ascent used in gradient based optimization methods. It is connected to the derivative of $f$ via ${\mderiv{\mat{Z}^\ast}{f} = \mvec{\mgrad{\mat{Z}}{f}}^\top}$.

\section{Derivative of the Singular Value Decomposition}
\label{sec:SVD-deriv}

In this section we derive the differentials of the singular value decomposition (SVD) of a matrix ${\A \in \mathbb{C}^{M \times N}}$. The SVD is defined as
\begin{equation}
    \label{eq:svd}
    \A = \V\, \S\, \W^\dagger \;,
\end{equation}
with (semi-)unitary matrices ${\V \in \mathbb{C}^{M \times K}}$ and ${\W \in \mathbb{C}^{N \times K}}$, \ie ${\V\, \V^\dagger = \W\, \W^\dagger = \mat{I}_K}$, and a diagonal matrix ${\S \in \mathbb{R}^{K \times K}}$ with positive entries ${\sigma_i\;; i=1,\ldots,K}$ called singular values. The goal is to derive expressions for the change in $\S$, $\V$, and $\W$ \wrt a change in $\A$.

Differentiating \cref{eq:svd} yields
\begin{equation}
    \label{eq:svd-diff}
    \D\A = \D\V\, \S\, \W^\dagger + \V\, \D\S\, \W^\dagger + \V\, \S\, \D\W^\dagger \;.
\end{equation}
By multiplying \cref{eq:svd-diff} from the left with $\V^\dagger$ and from the right with $\W$ we obtain
\begin{equation}
    \label{eq:dP}
    \D\P := \V^\dagger\, \D\A\, \W = \D\cV\, \S + \D\S + \S\, \D\cW^\dagger \;,
\end{equation}
with ${\D\cV := \V^\dagger\, \D\V}$ and ${\D\cW := \W^\dagger\, \D\W}$. Since $\V$ is (semi-)unitary, it holds
\begin{equation}
    \D(\V^\dagger\, \V) = 0 = \D\V^\dagger\, \V + \V^\dagger\, \D\V = \D\cV^\dagger + \D\cV \;,
\end{equation}
\ie $\D\cV$ is skew-Hermitian. It is convenient to decompose $\D\cV$ into ${\D\cV = \D\ctV + \D\cdV}$, where $\D\ctV$ is skew-Hermitian with zero diagonal and $\D\cdV$ is purely imaginary diagonal. The same considerations hold for $\D\cW$.

For the diagnoal elements of \cref{eq:dP} we obtain
\begin{equation}
    \D\P_{ii} = \D\cdV_{ii}\, \sigma_i + \D\sigma_i - \sigma_i\, \D\cdW_{ii} \;,
\end{equation}
and using the fact that the singular values (and their changes) are real and that $\D\cdV$ and $\D\cdW$ are purely imaginary, we obtain
\begin{align}
    \label{eq:dP-Re}
    \operatorname{Re}\lbrace \D\P_{ii} \rbrace &= \D\sigma_i \\
    \label{eq:dP-Im}
    \I \operatorname{Im}\lbrace \D\P_{ii} \rbrace &= (\D\cdV_{ii} - \D\cdW_{ii}) \sigma_i \;,
\end{align}
where the first equation describes the change in the singular values and the second equation gives the connection between $\D\cdV$ and $\D\cdW$.

From the off-diagonal elements of \cref{eq:dP} and it's complex conjugate transpose we obtain
\begin{align}
    \label{eq:dP-system}
    \begin{split}
        \D\P_{ij} &= \D\ctV_{ij}\, \sigma_j - \sigma_i\, \D\ctW_{ij} \\
        \D\P_{ji}^\ast &= -\sigma_i\, \D\ctV_{ij} + \D\ctW_{ij}\, \sigma_j \;.
    \end{split}
\end{align}
When $\sigma_i$ and $\sigma_j$ differ, this set of equations is solved by
\begin{align}
    \begin{split}
        (\sigma_j^2 - \sigma_i^2) \D\ctV_{ij} &= \D\P_{ij}\, \sigma_j + \sigma_i\, \D\P_{ji}^\ast \\
        (\sigma_j^2 - \sigma_i^2) \D\ctW_{ij} &= \sigma_i\, \D\P_{ij} + \D\P_{ji}^\ast\, \sigma_j \;.
    \end{split}
\end{align}
In the case of degenerate singular values, \cref{eq:dP-system} needs to be solved in a least squares fashion, yielding
\begin{align}
    \begin{split}
        4\sigma_i\, \D\ctV_{ij} &= \D\P_{ij} - \D\P_{ji}^\ast \\
        4\sigma_i\, \D\ctW_{ij} &= -\D\P_{ij} + \D\P_{ji}^\ast \;.
    \end{split}
\end{align}
Defining the matrices
\begin{equation}
    (\F_1)_{ij} = \begin{cases}
        0 & i = j \\
        \frac{\sigma_j}{\sigma_j^2 - \sigma_i^2} & \sigma_i \neq \sigma_j \\
        \frac{1}{4\sigma_i} & \sigma_i = \sigma_j
    \end{cases} \quad \text{and} \quad
    (\F_2)_{ij} = \begin{cases}
        0 & i = j \\
        \frac{\sigma_i}{\sigma_j^2 - \sigma_i^2} & \sigma_i \neq \sigma_j \\
        -\frac{1}{4\sigma_i} & \sigma_i = \sigma_j
    \end{cases}
\end{equation}
we can express the solutions to \cref{eq:dP-system} as
\begin{align}
    \label{eq:dcVW-tilde}
    \begin{split}
        \D\ctV &= \F_1 \odot \D\P + \F_2 \odot \D\P^\dagger \\
        \D\ctW &= \F_2 \odot \D\P + \F_1 \odot \D\P^\dagger \;.
    \end{split}
\end{align}
Note that ${\F_1^\top = -\F_2}$. Combining \cref{eq:dP-Im,eq:dcVW-tilde} we eventually arrive at
\begin{align}
    \label{eq:dcVW}
    \begin{split}
        \D\cV &= \F_1 \odot \D\P + \F_2 \odot \D\P^\dagger + \D\cdW + \frac{1}{2} (\D\P - \D\P^\dagger) \odot \S^{-1} \\
        \D\cW &= \F_2 \odot \D\P + \F_1 \odot \D\P^\dagger + \D\cdW \;,
    \end{split}
\end{align}
where we are still free to choose the imaginary diagonal matrix $\D\cdW$.

The last remaining task is to compute $\D\V$ from ${\D\cV = \V^\dagger\, \D\V}$ and doing the same for $\D\W$. To this extent, we fix a matrix $\V_\perp$ such that $[\V\; \V_\perp]$ is a square ${M \times M}$ unitary matrix. Then, $\D\V$ can be written as
\begin{equation}
    \D\V = \V\, \D\cV + \V_\perp\, \D\mat{K}^V \;,
\end{equation}
with an arbitrary ${(M-K) \times K}$ matrix ${\D\mat{K}^V}$. In order to fix ${\D\mat{K}^V}$ we left multiply \cref{eq:svd-diff} with $\V_\perp^\dagger$ which gives ${\V_\perp^\dagger\, \D\A = \D\mat{K}^V\, \S\, \W^\dagger}$ from which follows
\begin{align}
    \label{eq:dKVW}
    \begin{split}
        \D\mat{K}^V &= \V_\perp^\dagger\, \D\A\, \W\, \S^{-1} \\
        \D\mat{K}^W &= \W_\perp^\dagger\, \D\A^\dagger\, \V\, \S^{-1} \;,
    \end{split}
\end{align}
where the second equations follows from the same considerations for $\W$ instead of $\V$.

We now summarize \cref{eq:dP-Re,eq:dcVW,eq:dKVW} together with the equalities ${\V_\perp^\dagger\, \V_\perp = \mat{I}_M - \V\, \V^\dagger}$ and ${\W_\perp^\dagger\, \W_\perp = \mat{I}_N - \W\, \W^\dagger}$ and obtain the final result for the derivative of the SVD
\begin{align}
    \label{eq:dS}
    \D\S &= \mat{I}_K \odot \operatorname{Re}\lbrace \D\P \rbrace \\
    \label{eq:dV}
    \D\V &= \V \left( \F_1 \odot \D\P + \F_2 \odot \D\P^\dagger + \D\mat{D} + \frac{1}{2}(\D\P - \D\P^\dagger) \odot \S^{-1} \right) + (\mat{I}_M - \V\, \V^\dagger) \D\A\, \W\, \S^{-1} \\
    \label{eq:dW}
    \D\W &= \W \left( \F_2 \odot \D\P + \F_1 \odot \D\P^\dagger + \D\mat{D} \right) + (\mat{I}_N - \W\, \W^\dagger) \D\A^\dagger\, \V\, \S^{-1} \;,
\end{align}
with $\D\P$, $\F_1$, and $\F_2$ as defined above and an arbitrary purely imaginary diagonal matrix $\D\mat{D}$. Note that, if $\A$ is invertible, the last summands in \cref{eq:dV,eq:dW} vanish.

\section{Derivatives of the Wannier Gauge Matrices}
\label{sec:U-deriv}

The goal of this section is to find an expression for the derivatives $\mderiv{\X}{\U}$ and $\mderiv{\X^\ast}{\U}$, where $\X$ is the ${M \times J}$ matrix describing the OPFs and $\U$ is one of the ${N \times J}$ Wannier gauge matrices $\U^{\vec{k}}$. $\U$ is obtained from $\X$ by performing a SVD
\begin{align}
    \A\, \X &= \V\, \S\, \W^\dagger \nonumber \\
    \label{eq:U_def}
    \U &= \V\, \W^\dagger \;,
\end{align}
where $\A$ (again we ommit the wavevector $\vec{k}$) is the ${N \times M}$ overlap between single particle states and the trial projection functions. From \cref{eq:U_def} it follows
\begin{equation}
    \label{eq:dU_def}
    \D\U = \D\V\, \W^\dagger + \V\, \D\W^\dagger \;.
\end{equation}
We exploit the results of \cref{sec:SVD-deriv} in form of \cref{eq:dV,eq:dW}. Note the following replacements between \cref{sec:SVD-deriv} and this section:
\begin{align*}
    \A &\rightarrow \A\, \X \\
    \D\A &\rightarrow \A\, \D\X \\
    \D\P &\rightarrow \V^\dagger\, \A\, \D\X\, \W .
\end{align*}
Using the following identities
\begin{align}
    \label{eq:helpers}
    \begin{split}
        \mvec{\mat{A}\, \mat{B}\, \mat{C}} &= (\mat{C}^\top \otimes \mat{A}) \mvec{\mat{B}} \\
        \mvec{\mat{A} \odot \mat{B}} &= \diag{\mvec{\mat{A}}} \mvec{\mat{B}} \\
        \mvec{\mat{A}^\top} &= \mat{K}\, \mvec{\mat{A}} \;,
    \end{split}
\end{align}
where $\mat{K}$ is a permutation matrix of appropriate size, we compute $\mvec{\D\U}$ and find the following derivatives according to \cref{eq:mderiv_def} as the prefactors of $\mvec{\D\X}$ and $\mvec{\D\X^\ast}$, respectively:
\begin{align}
    \begin{split}
        \label{eq:deriv-U-X}
        \mderiv{\X}{\U} &= \left[ (\W^\ast \otimes \V) \diag{\mvec{\F}}\, (\W^\top \otimes V^\dagger) + (\W\, \S^{-1}\, \W^\dagger)^\top \otimes (\mat{I}_N - \V\, \V^\dagger) \right. \\
            &\quad \left. + (\mat{I}_J - \W\, \W^\dagger)^\top \otimes \V\, \S^{-1}\, \V^\dagger \right] \otimes (\mat{I}_J \otimes \A)
    \end{split}\\
    \begin{split}
        \label{eq:deriv-U-Xc}
        \mderiv{\X^\ast}{\U} &= - (\W^\ast \otimes \V) \diag{\mvec{\F}}\, \mat{K}\, (\W^\dagger \otimes \V^\top) \otimes (\mat{I}_J \otimes \A^\ast) \;,
    \end{split}
\end{align}
where we have introduced the matrix ${\F = \F_1 - \F_2 + \frac{1}{2}\S^{-1}}$ with elements
\begin{equation}
    \F_{ij} = \begin{cases}
        \frac{1}{2\sigma_i} & \sigma_i = \sigma_j \\
        \frac{\sigma_j - \sigma_i}{\sigma_j^2 - \sigma_i^2} & \sigma_i \neq \sigma_j
    \end{cases}\;.
\end{equation}
Note that the result is independent of the arbitrary imaginary diagonal matrix $\D\mat{D}$.

\section{Gradient of the Spread Functional}
\label{sec:spread-grad}

In order to minimize the spread $\Omega$ \wrt $\X$ using gradient based optimization algorithms we need to find the gradient $\mgrad{\X}{\Omega}$ which can be found by calculating 
\begin{equation}
    \mvec{\mgrad{\X}{\Omega}} = \mderiv{\X^\ast}{\Omega}^\top = \mderiv{\X}{\Omega}^\dagger \;.
\end{equation}
According to the chain rule \cref{eq:chain-rule} the derivative reads
\begin{equation}
    \mderiv{\X}{\Omega} = \sum\limits_{\vec{k}} \mderiv{\U^{\vec{k}}}{\Omega} \mderiv{\X}{\U^{\vec{k}}} + \mderiv{\U^{\vec{k}\ast}}{\Omega} \mderiv{\X}{\U^{\vec{k}\ast}} \;.
\end{equation}
Using the properties ${\mderiv{\U^{\vec{k}\ast}}{\Omega} = \mvec{\mgrad{\U^{\vec{k}}}{\Omega}}^\top}$ and ${\mderiv{\U^{\vec{k}}}{\Omega} = \mderiv{\U^{\vec{k}\ast}}{\Omega}^\ast}$ we find the following expression for calculating the gradient
\begin{equation}
    \mvec{\mgrad{\X}{\Omega}} = \sum\limits_{\vec{k}} \mderiv{\X}{\U^{\vec{k}}}^\dagger \mvec{\mgrad{\U^{\vec{k}}}{\Omega}} + \mderiv{\X^\ast}{\U^{\vec{k}}}^\top \mvec{\mgrad{\U^{\vec{k}}}{\Omega}}^\ast \;.
\end{equation}
We insert the results of \cref{sec:U-deriv} in form of \cref{eq:deriv-U-X,eq:deriv-U-Xc} and again employ the identities \cref{eq:helpers} to find the final result for the gradient of the spread functional
\begin{align}
    \label{eq:spread-grad}
    \mgrad{\X}{\Omega} &= \sum\limits_{\vec{k}} \A^\dagger \left\lbrace \V \left( \F \odot \left[ \V^\dagger\, \mgrad{\U}{\Omega}\, \W \right] - \text{H.c.} \right) \W^\dagger \right. \nonumber \\
        &\quad \left. + (\mat{I}_N - \V\, \V^\dagger) \mgrad{\U}{\Omega} \W\, \S^{-1}\, \W^\dagger + \V\, \S^{-1}\, \V^\dagger \mgrad{\U}{\Omega} (\mat{I}_J - \W\, \W^\dagger) \right\rbrace \;.
\end{align}
Note that all matrices on the right hand side of \cref{eq:spread-grad} are explicitly $\vec{k}$-dependent. 

\bibliography{bibliography}

%% file: math_macros.tex
\newcommand{\I}{\mathrm{i}\mkern1mu}
\newcommand{\E}{\mathrm{e}\mkern1mu}

\newcommand{\D}{\mkern1mu\mathrm{d}}
\renewcommand{\vec}[1]{\bm{#1}}
\newcommand{\mat}[1]{\bm{\mathrm{#1}}}

\usepackage{xparse}
\NewDocumentCommand{\mDscal}{m O{} m O{}}{\ensuremath{( \mathcal{D}_{\scriptstyle \mat{#3}#4} #1#2)}}
\NewDocumentCommand{\mDmat}{m O{} m O{}}{\ensuremath{( \mathcal{D}_{\scriptstyle \mat{#3}#4} \mat{#1}#2)}}
\NewDocumentCommand{\mGrad}{m O{} m O{}}{\ensuremath{( \nabla_{\scriptstyle \mat{#3}#4} #1#2)}}

%% file: acronyms.tex
\newacronym{dft}{DFT}{density-functional theory}
\newacronym{wf}{WF}{Wannier function}
\newacronym{mlwf}{MLWF}{maximally localized Wannier function}
\newacronym{opf}{OPF}{optimized projection function}
\newacronym{svd}{SVD}{singular value decomposition}
\newacronym{lapw}{(L)APW}{(linearized) augmented plane wave}
\newacronym{lo}{LO}{local orbital}
\newacronym{lbfgs}{L-BFGS}{limited-memory Broyden--Fletcher--Goldfarb--Shanno}
\newacronym{pao}{PAO}{pseudo-atomic orbital}
\newacronym{scdm}{SCDM}{selected columns of the density matrix}

%% file: tables/spread_isolated.tex
\begin{tabularx}{\linewidth}{X D{.}{}{1} D{.}{}{1} D{.}{}{1} d d d d}
     \toprule
     &  \multicolumn{1}{c}{$J$} & 
        \multicolumn{2}{c}{$M$} &
        \multicolumn{1}{c}{$\gamma$} &
        \multicolumn{2}{c}{$\Omega^\textrm{OPF}$} &
        \multicolumn{1}{c}{$\Omega^\textrm{MLWF}$} \\
     & & \multicolumn{1}{c}{home} & \multicolumn{1}{c}{+nn} &
       & \multicolumn{1}{c}{home} & \multicolumn{1}{c}{+nn} \\
     \midrule
     c-Si               &  4    &  8    &  20   & 0.39  &   6.569   &   6.513   &   6.512   \\ %
     Si-20              & 40    & 80    & 144   & 0.38  & 102.343   & 101.114   & 101.011   \\ %
     GaAs               &  4    & 11    &  23   & 0.40  &   7.272   &   7.206   &   7.206   \\ %
     SiO\sub{2}         &  8    & 18    &  34   & 0.57  &   9.119   &   9.119   &   9.119   \\ %
     Cr\sub{2}O\sub{3}  & 12    & 38    &  74   & 0.84  &  27.105   &  27.085   &  27.057   \\ %
     BaSnO\sub{3}       &  9    & 25    &  73   & 0.63  &  12.145   &  12.142   &  12.141   \\ %
     NaCl               &  3    &  4    &  19   & 0.28  &   3.900  &   3.900   &   3.900   \\ %
     \bottomrule
\end{tabularx}

%% file: tables/spread_entangled.tex
\begin{tabularx}{\linewidth}{X X X D{.}{}{1} D{.}{}{1} D{.}{.}{3} D{.}{.}{3} D{.}{.}{3}}
     \toprule
     &  \multicolumn{1}{c}{$E^{\rm wind}$} &
     &
        \multicolumn{1}{c}{$J$} & 
        \multicolumn{1}{c}{$M$} &
        \multicolumn{1}{c}{$\gamma$} &
        \multicolumn{1}{c}{$\Omega^{\textrm{OPF}}$} & 
        \multicolumn{1}{c}{$\Omega^{\textrm{MLWF}}$}\\
     \midrule
     Si-20          & 20 (25)   & home   & 160   & 188   & 0.32  & 1094.7	   & 696.4 \\ %
                    &          & +nn    & 		& 339	&		&  848.3    & 696.4	\\ %
     SiO\sub{2}     & 40 (50)   & home   &  79   & 112   & 0.17  &  446.8	   & 182.0 \\ %
                    &          & +nn	& 		& 192	& 	    &  346.3    & 181.7	\\ %
     Zn             & 70 (85)   & home   &  64   &  92   & 0.55  &  183.4    &  105.0 \\ %
                    &          & +nn	& 		& 506	& 		&   139.3    &  106.0	\\ %
     \bottomrule
\end{tabularx}

%% file: tables/spread_self-projection.tex
\begin{tabularx}{\linewidth}{X X D{.}{.}{3} D{.}{.}{3} D{.}{}{1} D{.}{.}{3}}
     \toprule
     &  &   \multicolumn{1}{c}{$\Omega^{\textrm{OPF}}$} & 
            \multicolumn{1}{c}{$\Omega^{\textrm{OPF+sp}}$} & 
            \multicolumn{1}{c}{$\Delta\Omega^{\textrm{sp}}$} & 
            \multicolumn{1}{c}{$\Omega^{\textrm{MLWF}}$}\\
     \midrule
     Si-20          & home  & 1094.7	   & 842.8     & 23\%  & 696.4 \\ %
                    & +nn   &  848.3    & 758.8		& 11\%	& 696.4	\\ %
     SiO\sub{2}     & home  &  446.8	   & 386.4     & 14\%  & 182.0 \\ %
                    & +nn   &  346.3    & 266.0		& 23\%	& 181.7	\\ %
     Zn             & home  &  183.4    & 175.1     &  5\%  & 105.0 \\ %
                    & +nn   &  139.3    & 133.9     &  4\%  & 106.0	\\ %
     \bottomrule
\end{tabularx}